# Comparison Between International Space Station and Airplane Flying

M. N. Tarabishy, Ph.D.

goodsamt@gmail.com

**Abstract:**

Inertial forces are not intuitive, therefore, interesting examples are great help for learners. In this paper, we examine simple models for International Space Station (ISS) and airplane flying and identify the inertial forces involved and their effect. In both cases there are centrifugal forces but in the case of the ISS, the centrifugal force is the only major force opposing the gravitational attraction that is pulling it towards the Earth at 0.885 the gravity at the surface of the Earth. While in the case of an airplane, it is a minor component amounting to less than 1% in the best case of flying along the equator west to east. Other inertial forces including the Coriolis force affect the airplane. Another thing we see is that flying with the Earth direction west to east gives a markedly higher forces than flying in the opposite direction.

The maintenance of the ISS orientation with respect to Earth requires rotation of ISS around its center of gravity (cg) at the same rate as it revolves around the Earth (4 degrees per minute) while for an airplane, maintaining level flight at a constant speed and height does not require direct intervention because the amount of adjustment is so small and gradual that it is handled by the pilot system as a part of the disturbances that affects the airplane.

## 1. Introduction:

When an aircraft cruises at constant altitude with constant speed, then clearly, enough lift has to be generated to counteract the gravitational pull or the weight of the airplane and that is done almost entirely by the aerodynamic forces.

The other issue is that the airplane must keep parallel to the Earth as it curves underneath it.

The argument used by flat earthers is that if the Earth was a sphere, then the pilot needs to adjust the nose of the plane downwards continuously to keep up with Earth curvature while flying at constant height, but since pilots don't do that, therefore, the Earth must be flat.

At the surface, this argument seems to make sense, but a simple analysis would invalidate such an argument. In the following sections, we will look at the example of the ISS motion and compare that to an example of an airplane flying.

## 2. The ISS Motion:

The ISS orbits around the earth at an average height of 400 km above the surface and at a speed of 7778 m/s, so, it completes one revolution in about 90 minutes.

If we analyze its motion with Newton's second law.

$$F = m \cdot a \tag{1}$$

F: The total forces affecting the ISS.

m: The mass of the ISS = 420E3 kg.

a: Acceleration of the center of gravity (cg) of the ISS.

The force is given according to Newton's gravitational law:

$$F = \frac{G \cdot m \cdot M}{r^2} = m \, g_{ISS} \tag{2}$$

r: The distance from the center of the Earth to the ISS = (6370 + 400)E3 = 6770E3 m.

G: gravitational constant = 6.67E-11 N.m² / kg².

M: Earth mass = 6E24 kg.

$g_{ISS}$: The gravitational constant at the ISS height.

However, the gravitational acceleration at the height of the ISS is related to g, the gravitational acceleration at the surface of the Earth by:

$$\frac{g}{g_{ISS}} = \frac{r_{ISS}^2}{r^2} \tag{3}$$

g: Gravitational acceleration at the surface of the planet = 9.82 m/s².

r: Earth radius = 6370E3 m.

Therefore, we find that the gravitational acceleration affecting the ISS is $g_{ISS}$ = 0.885 g.

The motion of the satellite is planar so we can use polar coordinates to describe its motion:

$$(\ddot{r} - r \dot{\theta}^2) \hat{r} + (r \ddot{\theta} + 2 \dot{r} \dot{\theta}) \hat{\theta} = (F/m) \hat{r} \tag{4}$$

For circular motion at constant angular velocity, the θ̂ component is zero, and the equation of motion becomes:

$$r\dot{\theta}^2 - 0.885\,g = 0 \qquad (5)$$

$\dot{\theta}$ : The angular velocity of the ISS (rad/s).

It is simply stating that the inertial force represented by the centrifugal force is equal and opposite to the gravitational attraction force, and this balance is depicted schematically by the red arrows in figure 1. So, the ISS keeps falling towards the Earth, but its speed keeps it in orbit.

To keep its belly towards the Earth, the ISS rotates around its cg at 4 degrees / minute, otherwise, it will remain oriented like the shadow depicted in figure 1.

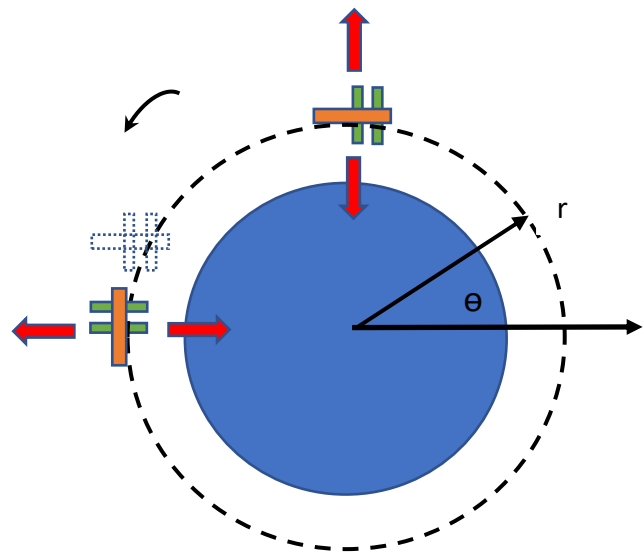

Figure 1. The ISS revolution around the Earth and rotation around its cg

### 3. Airplane Motion Analysis:

While the forces acting on the ISS are mainly gravity "weight" which is balanced by the centrifugal force, for the airplane, it is a bit different.

The forces acting on the airplane are the weight, the lift, thrust, drag, and inertial forces. The weight of the airplane acts at the cg and points in the radial direction towards the center of the Earth, and this force is countered by the lift force to keep the airplane aloft.

In [1], a simplified analysis is used by assuming that the path of the airplane is on a great circle and that two-dimensional approach is sufficient, however, three-dimensional analysis gives us interesting details that are not available in the simplified 2-D analysis, so, we will use spherical coordinates for analysis as shown in figure 2.

The first coordinate is radial distance r from the center of the Earth to the airplane, and its direction is indicated by the direction unit vector r̂ (locally, it will be the up direction). The second coordinate is the inclination or zenith angle ∅ which is the angle that r makes with the positive z axis. This angle is related to the latitude angle λ by:

$$\emptyset = 90 - \lambda \qquad (6)$$

The third coordinate is the polar or azimuth angle θ which corresponds to the longitude angle as shown in figure 2.

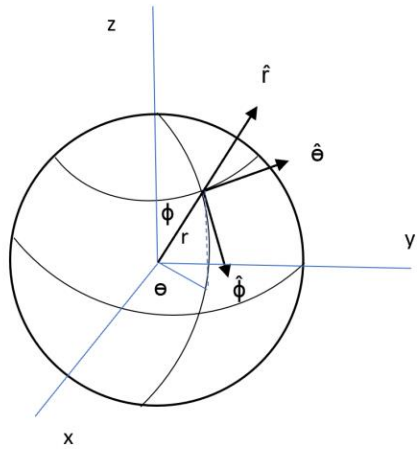

Figure 2. The radial distance r, zenith ∅ and azimuth θ angles are shown

The equations of motion given by Newton's second law in spherical coordinates are, [2]:

$$[\ddot{r} - r\dot{\emptyset}^2 - r\dot{\theta}^2 \sin^2(\emptyset)]\ \hat{r} + \qquad (7)$$
$$[r\ddot{\theta}\sin(\emptyset) + 2\dot{r}\dot{\theta}\sin(\emptyset) + 2r\dot{\theta}\dot{\emptyset}\cos(\emptyset)]\hat{\theta} +$$
$$[r\ddot{\emptyset} + 2\dot{r}\dot{\emptyset} - r\dot{\theta}^2\sin(\emptyset)\cos(\emptyset)]\ \hat{\emptyset}$$
$$= \frac{1}{m}\{[(L-W)]\hat{r} + [(T-D)_{east}]\hat{\theta} + [(T-D)_{south}]\hat{\emptyset}\}$$

Where:

r: Radius from the Earth center (m).

θ: Polar "Azimuth" angle (rad).

∅: Inclination "Zenith" angle (rad).

θ̇, ∅̇: Angular speeds resulting from going west-east and north-south respectively (rad/s).

r̂, θ̂, ø̂: Unit direction vectors.

r̈, θ̈, ø̈: Linear (m/s²) and angular accelerations (rad/s²).

L: The lift force (N).

W: The weight of the plane (N)

m: The mass of the plane (kg).

T: Thrust force (N).

D: Drag force (N).

For constant speed at constant height, and taking the rotation of the Earth into consideration, the equations of motion become:

$$[-r\dot{\emptyset}^2 - r\dot{\Omega}^2 \sin^2(\emptyset)]\,\hat{r} + \\ [r\ddot{\theta}\sin(\emptyset) + 2r\dot{\Omega}\dot{\emptyset}\cos(\emptyset)]\hat{\theta} + \\ [r\ddot{\emptyset} - r\dot{\Omega}^2\sin(\emptyset)\cos(\emptyset)]\hat{\emptyset} \\ = \frac{1}{m}\{[(L-W)]\hat{r} + [(T-D)_{lat}]\hat{\theta} + [(T-D)_{long}]\hat{\emptyset}\}$$ (8)

Where:

$$\Omega = \dot{\theta} + \omega$$ (9)

ω: Angular speed of the Earth (rad/s).

The radial direction r̂ is the main direction where the two acceleration terms give two centrifugal forces that are added to the lift force.

The θ̂ component is along the latitude and it is positive in the east direction. It mainly gives the Coriolis force which has the form ( -2m r θ̇ ø̇ cos(ø) ). The magnitude of this force depends on the latitude angle λ, with a maximum at the poles to zero at the equator.

The ø̂ component is along the longitude and positive direction is south. It mainly gives a centrifugal force projection that help the thrust in this case. This force is also a function of latitude angle λ. It is worth noting that the vectors ø̂, r̂, and ẑ are coplanar.

If the motion of the airplane is west to east, then, the resulting centrifugal force is larger due to the addition of rotation from the motion to Earth's rotation. On the other hand, if the motion is east to west, then, the two angular motions are subtracted.

To put these terms into perspective, we will take specific examples.

### 4. Example 1:

A plane that is flying at 10,000 m (about 33,000 ft) at a speed of 223.5 m/s (500 mph). The lift force is the result of mainly the interaction between the wings and the streaming air at a suitable angle of attack. This lift overcomes the weight force while the contribution of the centrifugal force opposing the weight is fairly small as we will see.

The thrust generated by the engines overcomes the drag force opposing the motion of the airplane, and when the forces are balanced, the aircraft cruises at constant speed.

If the plane were flying around the equator in the west to east direction, then, its trajectory is on a great circle and the airplane angular velocity around the Earth is 3.5E-5 rad/s. At such speed it would complete a revolution around the Earth in 50 hours.

The Earth angular velocity is 7.3E-5 rad/s, therefore, the Earth has double the angular speed, but the atmosphere moves with the Earth, so the aircraft has to fly with respect to the Earth/atmosphere to get from one point to another. The centrifugal force resulting from the Earth spinning affects the airplane.

The centripetal acceleration resulting from the airplane motion and the earth spin amounts to 0.074 m/s$^2$, a very small value compared to the gravitational acceleration g = 9.82 m/s$^2$. If we multiply this acceleration with the mass of the airplane, we will get the value of the centrifugal force.

For an airliner of about 500 tons, it is 37000 N, making the airplane appears to be lighter by about 4 tons.

However, if we reverse the direction of travel to one from east to west, then, the centrifugal force is 4534 N. A much smaller amount because the angular velocity of travel is now subtracted from the angular velocity of Earth.

On its own, without considering the Earth's rotation effect, the centrifugal force is 3915 N.

### 5. Example 2:

We will take a more general example of the same airplane in the previous example flying at the same height and speed also, but it goes from Chicago to L.A. on a great circle path which is shown in figure 3.

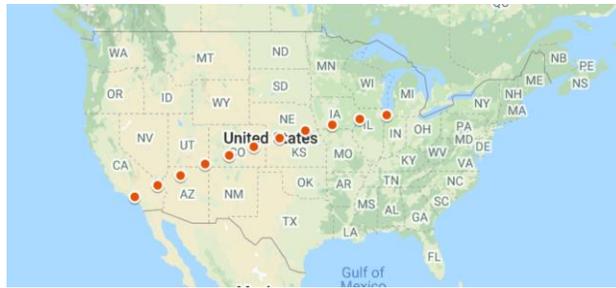

Figure 3. The great circle path from Chicago to L.A.

We have calculated the forces that affect the airplane at each one of these 11 points (ignoring climbing and descending parts). We numbered the points from Chicago (1) to L.A. (11) and we plotted the results from point 3 to point 9 because we used the first two points to calculate the accelerations. The inertial forces are plotted in figures 4, 5, and 6, for three cases:

The first case:  Ignoring the Earth spinning effect and considering only the motion of the airplane (without earth effect)

The second case:  Adding the Earth spinning to the motion (east to west).

The third case:  This time we fly from L.A. to Chicago, and we consider Earth spinning in addition to flying motion (west to east).

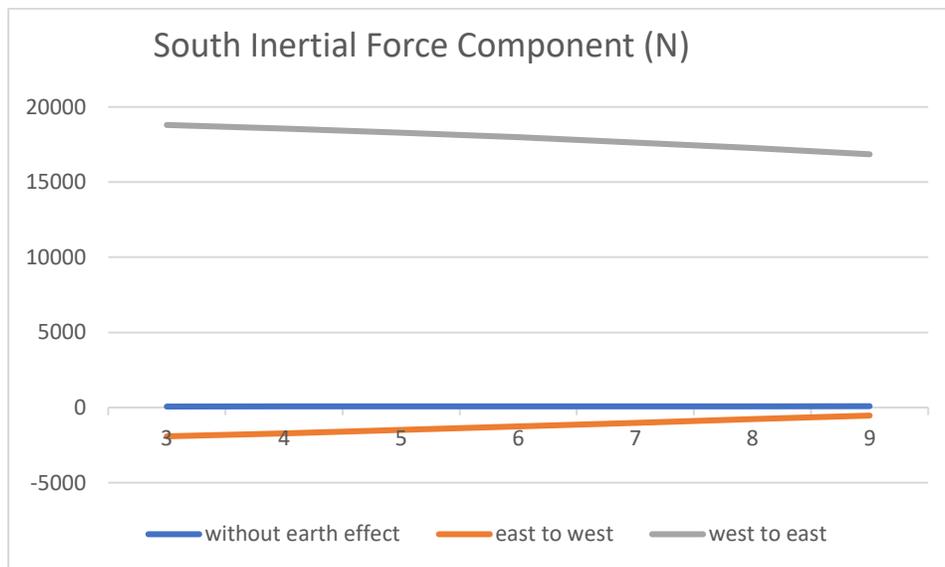

Figure 4. The south inertial force components

In figure 4, we see that there is a marked difference between going east to west and the opposite direction.

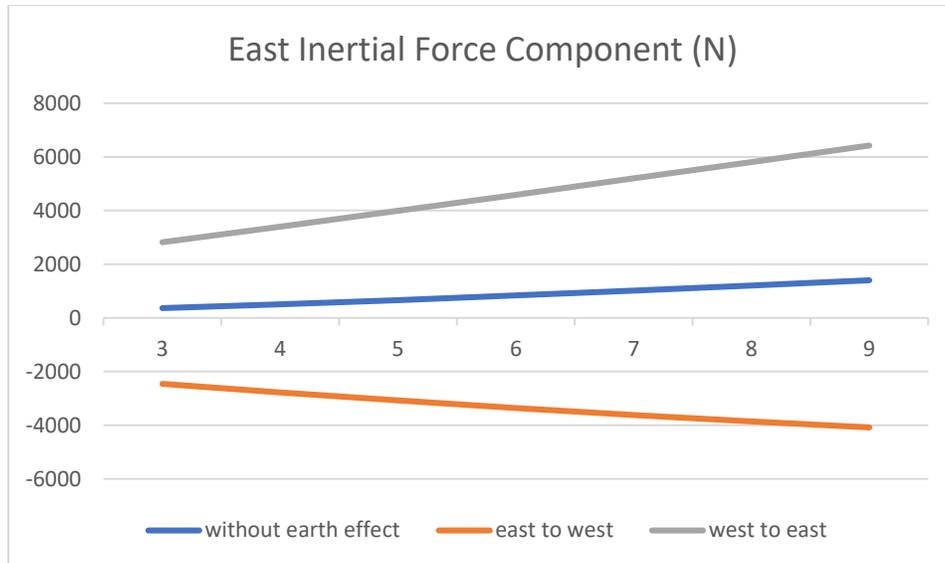

Figure 5. The east inertial force components

In figure 5, we see a depiction of mainly the Coriolis force for the three cases. The Coriolis force (with motion in a rotating Earth) in the northern hemisphere is to the right of the motion, so going from Chicago in the southwest direction, the force direction is in westerly direction, giving it a negative sign as we can see with orange curve. Going northeastern from L.A., the force has a easterly direction and therefore, it has a positive sign shown in the gray curve. Without considering the Earth spinning, this force is the smallest.

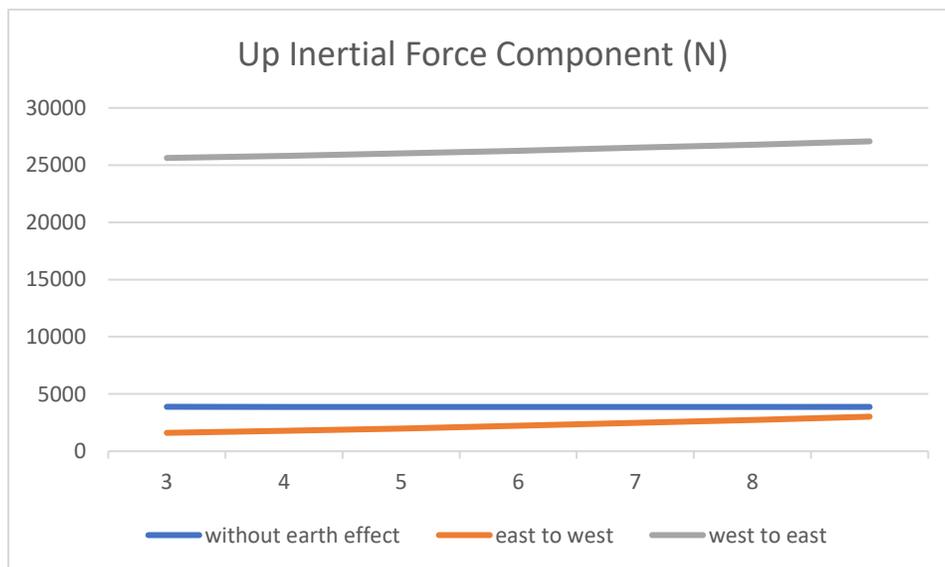

Figure 6. The up inertial force components

In figure 6, we see that all these components are positive, but there is a marked difference in the magnitude of centrifugal force between going east to west and the much bigger value for going west to east.

**6. Level Flight:**

To achieve level flight, the forces as well as the torques around the cg should be balanced. The main forces on the airplane are the weight acting at the cg (the blue dot in figure 7), and the lift force balancing it. Then, there is the thrust force overcoming the drag force.

The lift force is divided into two main parts. The wings lift force Lw acting at center of pressure of the wings is usually located aft the cg and it creates a torque around the cg. This torque is balanced by the torque of the lift generated at the tail Lt as shown in figure 7.

When the airplane is not maneuvering, the rotation around the cg should be zero, and that means the sum of moments of wings lift Lw, and tail lift Lt around the cg should be zero and the airplane is said to be "trimmed".

To keep its level cruising, the airplane has to rotate 0.12 degrees / minute, an amount that is imperceptible and is pretty much a perturbation that is handled by the pilot system that keeps the airplane trimmed. Therefore, it does appear as if the Earth is flat and not spherical due to the very gradual curvature.

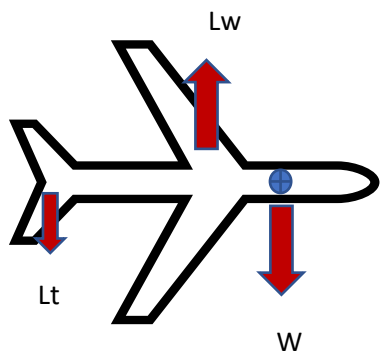

Figure 7. The moments of forces balance around the cg (blue)

**7. Discussion and Conclusions:**

To perform analysis, it is important to choose the appropriate reference frame. We have used three different frames. The first frame is rigidly attached to Earth and has its origin at the center of the Earth. Such a frame is used when we ignore the spinning of the planet. The second frame has its origin at the center of the Earth but doesn't spin with it. The third frame is the "local" which has its axis along the directions (South-East-Up). In this frame we experience everyday events and we used it to represent the forces with the corresponding components: $\hat{\phi}$ - $\hat{\theta}$ - $\hat{r}$.

In the analysis of the ISS, a simple 2-D model is a good model for its motion where the gravitational force (weight) is balanced by the centrifugal inertial force.

The attractive force on the ISS is 0.885 of its value at the surface of the Earth, therefore, the inertial centrifugal force should be large enough to balance the weight and that is why it has such a high velocity of 7778 m/s.

For the airplane, we used 3-D model to enable us to consider more things like Coriolis inertial forces, however, the first example of flying along the equator is very similar to the ISS except for the fact that the weight is, for all intents and purposes, is opposed by the lift force. The centrifugal force is really small in comparison. For an airliner of 500 tons, the weight is about 5 million Newtons, whereas the opposing centrifugal force at its best flying west to east is 37000 N amounting to less than 1% of the lift force needed, but it makes the craft appear lighter by about 3700 kg. If we use $ 4.00 / kg as estimate for air cargo cost [3], then, this will translate to about $ 15,000 in savings. It is worth noting that flying in the opposite direction results in a much smaller force of 4534 N.

In the second example, we had three cases. The first case is going from Chicago to L.A. on the great circle path without considering the Earth spinning, we notice that all components are small, with the centrifugal force of a bit lower than 4000 N.

In the second case, when we considered the Earth spinning, the Coriolis forces increased a bit while the centrifugal force is at a max of about 3000 N which is less than the first case.

The third case is when we went in the opposite direction from L.A. to Chicago (west to east). All force components are larger than the previous two cases. The Coriolis force is over 6000 N, and the centrifugal force in the up direction exceeded 27000 N, but it is smaller than the case of the equator flying.

We see that there is a clear difference between travelling east to west or west to east. However, the more significant factor is the Jet Streams that move the air towards the east giving a good boost to airplanes travelling west to east and opposing flights in the opposite direction [4]. The above discussion might be used to examine the benefit of west to east runway orientation (if we didn't consider that the primary consideration for orienting runways is the prevailing wind direction).

To keep the belly of the ISS pointing towards the Earth, the ISS must rotate around its cg by 4 degrees / minute in addition to its motion around the Earth.

For maintaining level flying of our airplane, the required rotation about its cg is 0.12 degrees / minute, an amount that is imperceptible, and since the pilot system usually handles disturbances that are much larger, then, this required rotation will fall within the adjustments that are done automatically to maintain level flying at constant speed, therefore, no need for constant adjustment of the airplane nose.

While the centrifugal and Coriolis inertial forces are small compared to the airflow forces in the case of flying and might be ignored from an engineering point of view, however, they are important from physics and pedagogical point of view.